\documentclass[12pt,preprint]{aastex}

\shorttitle{Disks with ATCA}
\shortauthors{Wilner et al.}

\begin{document}

\slugcomment{accepted by ApJ: June 18, 2003}

\title{Disks around the Young Stars TW Hya and HD~100546 
Imaged at 3.4 Millimeters with the Australia Telescope Compact Array}

\author{D.J. Wilner, T.L. Bourke}
\email{dwilner, tbourke@cfa.harvard.edu}
\affil{Harvard-Smithsonian Center for Astrophysics, 
60 Garden Street, Cambridge, MA 02138}

\author{C.M. Wright}
\email{wright@ph.adfa.edu.au}
\affil{School of Physics, University College, ADFA, UNSW,
Canberra ACT 2600, Australia}

\author{J. K. J{\o}rgensen, E.F. van Dishoeck}
\email{joergensen, ewine@strw.leidenuniv.nl}
\affil{Leiden Observatory, PO Box 9513, 2300 RA Leiden, The Netherlands}

\author{T. Wong}
\email{Tony.Wong@csiro.au}
\affil{CSIRO Australia Telescope National Facility, P.O. Box 76, 
Epping NSW 1710, Australia}

\begin{abstract}
We present observations of the young stars TW~Hya and HD~100546 
made with the Australia Telescope Compact Array at 89~GHz with 
$\sim2''$ resolution and $\sim3$~mJy continuum sensitivity.
Compact thermal dust continuum emission is detected from disks 
surrounding both stars. HD~100546 also shows hints of extended 
emission, presumably a residual protostellar envelope, 
which is also visible in scattered light at optical wavelengths.
For TW Hya, HCO$^+$ J=1--0 line emission from the circumstellar disk 
is detected and spatially resolved. The observed size and intensity
are in good agreement with model calculations based on an irradiated
disk with substantial depletions derived previously from single dish
observations of higher-J HCO$^+$ transitions. 
\end{abstract}

\keywords{circumstellar matter ---
planetary systems: protoplanetary disks ---
stars: individual (TW Hya, HD~100546)} 

\section{Introduction}

Many young stars exhibit emission from circumstellar dust particles 
distributed in disks with properties similar to the early Solar System,
and much activity is currently devoted to characterizing the physical 
properties of these disks to extract information on planet formation.
Observations at millimeter wavelengths are especially important 
because the disk material beyond a few stellar radii is at temperatures 
from a few hundreds to a few tens of degrees, and the physical 
and chemical conditions can be probed in detail in this part of the 
spectrum (see reviews by Beckwith \& Sargent 1996, 
Wilner \& Lay 2000, Langer et al. 2000).

The recent upgrade of the Australia Telescope Compact Array (ATCA)
with receivers for the 3~mm atmospheric window provides a new 
opportunity for high resolution imaging of protoplanetary disks 
located in the southern sky that are difficult or impossible to observe 
with the millimeter arrays located in the northern hemisphere.
The ATCA millimeter-wave upgrade is in progress, and the completed
facility will ultimately rival the northern interferometers in 
sensitivity and angular resolution (see Wong \& Melatos 2002 for
a more complete description). 
Now, three of the array antennas are equipped with 3~mm band InP~MMIC 
receivers, and an interim local oscillator system allows for tuning in 
two ranges between 85 and 91 GHz.
We present interferometric observations of dust continuum and 
HCO$^+$ J=1--0 line emission at 89 GHz from two southern disk targets: 
(1) TW Hya, the closest known classical T Tauri star,
and (2) HD~100546, a nearby Herbig Be star whose infrared spectrum
shows crystalline silicates, indicative of comet-like dust. 

TW Hya has received a lot of recent observational attention, largely
on account of its close proximity. At a distance of $56\pm7$ pc (Hipparcos), 
TW Hya is almost three times closer than the classical T Tauri stars 
associated with nearby dark clouds like Taurus and Ophiucus, which makes
it an especially attractive target for observations with high angular 
resolution. At an age of $\sim10$ Myr, the TW Hya disk appears to be 
substantially evolved, with indications of significant grain growth.
TW Hya is isolated from any molecular cloud but retains a face-on 
molecular disk visible in scattered light (Krist et al. 2000, 
Trilling et al. 2001, Weinberger et al. 2002) that extends to a radius 
of at least $3\farcs5$ (200 AU). 
The inner disk of TW Hya has been resolved in dust emission by the VLA
at 7~mm (Wilner et al. 2000), and its spectral energy distribution
has been modeled extensively (Calvet et al. 2001). The available data
are well fitted by an irradiated accretion disk with a developing gap of
radius $\sim4$~AU, interior to which dust coagulation and settling have
rendered the disk optically thin.

The isolation of the TW Hya system from any surrounding molecular cloud 
material facilitates the study of its physical and chemical structure. 
The TW Hya disk has been detected in a suite of molecular species
(CO, HCN, CN, and HCO$^+$) at submillimeter wavelengths using single dish 
telescopes (Kastner et al. 1997, van Zadelhoff et al. 2001).
More recent, deeper, observations have also detected emission from
H$^{13}$CO$^{+}$ and DCO$^{+}$, which indicate deuterium fractionation 
in the disk similar to cold cores and pristine cometary material
(van Dishoeck et al. 2003). So far, no observations have been made 
that spatially resolve the molecular gas surrounding TW Hya. 
Such spatially resolved observations are important to verify underlying 
constructs of the models used to interpret the single dish data, since the 
physical conditions are subject to strong radial and vertical gradients, 
which result in considerable chemical complexity. 

HD~100546, among the nearest Herbig Ae/Be stars at $103\pm6$ pc (Hipparcos),
appears nearby the dark cloud DC296.2-7.9 (Hu, The \& de Winter 1989) 
and shows a disk-like scattered light distribution 
(Pantin, Waelkens \& Lagage 2000,
Augereau et al. 2001, Grady et al. 2001) of substantial size, 
about $8''$ (8000~AU), as well as 
strong millimeter emission from dust (Henning et al. 1994, 1998).
This relatively isolated system is thought to have an age of $\sim10$ Myr
(van den Ancker et al. 1997).
However, analysis of the full spectral energy distribution suggests 
the presence of an extended envelope (Henning et al. 1994).
Mid-infrared spectroscopy from ISO shows remarkably strong crystalline 
silicate bands, similar to those observed in comet Hale-Bopp 
(Malfait et al. 1998) and indicative of unusual and substantial 
processing of the dust within the disk (Bouwman et al. 2003).
Millimeter interferometry has the potential to show directly the
presence of a disk component that contains the bulk of the system mass. 

\section{Observations}

We observed TW~Hya and HD~100456 at 89 GHz (3.4~mm) with the ATCA 
during the 2002 austral winter using two compact configurations
of three 22 meter diameter antennas equipped with millimeter receivers.
Table 1 summarizes the observational parameters. 

The ATCA observations provided 6 independent east-west baselines with
lengths ranging from 22 to 230 meters, resulting in $\sim2''$ resolution.
For each observation, the target pointing center was set to be 
$5''$ west of the star position. 
The digital correlator was configured for two dual polarization bands:
(1) a wide band with 33 channels spanning the maximum bandwidth of 128 MHz 
for continuum sensitivity, and (2) a narrow band with 256 channels over 
16 MHz to provide high frequency resolution on the HCO$^+$ J=1--0 line, 
which is one of the few potentially detectable spectral lines accessible 
to the interim system. The sky signal is mixed with a Gunn oscillator 
locked at 80505.5 MHz and passed through an X-band (8000-10800 MHz) 
filter and splitter module, 
which results in an accessible frequency range of 88.506 to 91.305 GHz.
A new local oscillator system under construction will allow coverage 
of the frequency range 85 to 105 GHz, with a possible extension to 115 GHz. 

All calibration and imaging was performed with the MIRIAD software.
Complex gains were derived from frequent observations of nearby quasars.
Because the narrow band provided low signal-to-noise ratios on the quasars, 
a phase offset between the wide band and narrow band 
was determined from short observations of a strong source on each day,
either J0423-012 or J0538-440. These short observations of strong sources
were also used to determine the bandpass response. 
For TW Hya, additional weaker quasars located close to the star 
were included in the observing sequence to provide an empirical check 
on the atmospheric seeing and effectiveness of the phase calibration.
Typical system temperatures were 300 to 400 K (SSB), with higher values
at low elevations.
The flux densities were set with reference to the planet Mars; 
the scatter in derived fluxes on consecutive days suggests uncertainties 
of approximately 20\%. 

Images were made of the sum of the two linear polarizations using 
natural weighting to obtain best sensitivity. All images were 
cleaned to a cutoff of twice the r.m.s. noise level.
Because Doppler tracking was not applied during the observations, the 
correspondence between individual frequency channels and velocity changes
with time during the course of the observations. Therefore, the spectral data 
were imaged in resampled velocity bins of 0.5~km~s$^{-1}$ width, coarse
compared to these changes, rather than individual frequency channels. 

\section{Results}

\subsection{TW Hya}

\begin{figure}
\epsscale{0.9}
\rotatebox{-90}{
\plotone{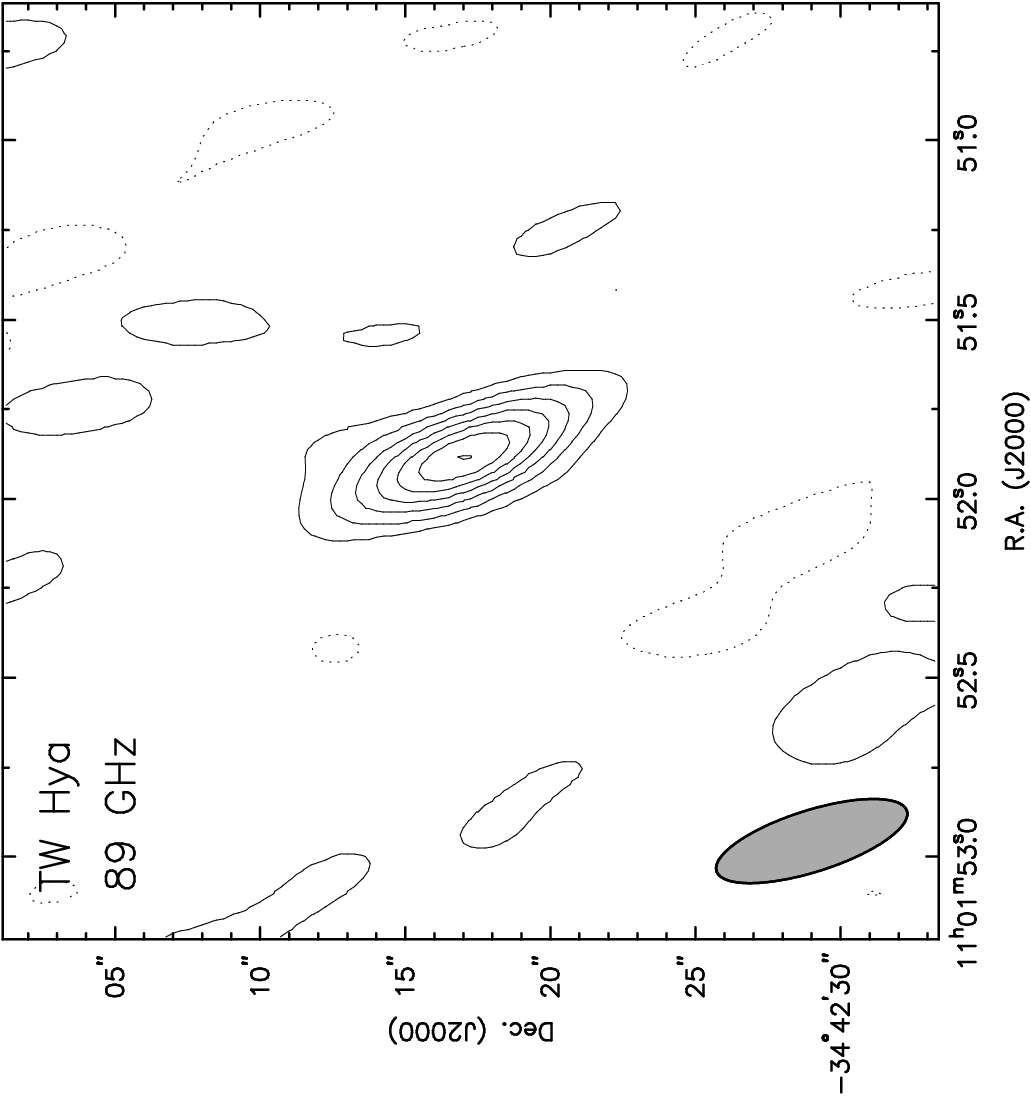}}
\caption{
TW Hya continuum emission at 89 GHz.
The contour levels are $\pm2,4,6,...\times2.5$~mJy.
Negative contours are dotted.
The ellipse in the lower left corner shows the 
$6\farcs9\times2\farcs1$ p.a. $18^{\circ}$ synthesized beam. 
}
\label{fig:tw_cont}
\end{figure}

Figure~\ref{fig:tw_cont} shows the 89 GHz continuum emission  
detected from TW~Hya. 
The position is coincident with the optical star and radio detections 
at longer wavelengths. 
A Gaussian fit to the visibilities gives a flux of $41\pm4$~mJy
(random error only) and an apparent size consistent with an unresolved
source slightly broadened by the phase noise on the longer baselines.
This Gaussian fit is also compatible with the extended disk seen
in scattered light, since the dust emission from the circumstellar disk 
is strongly centrally peaked by the combination of low opacity 
and increasing column density and temperature towards the star
(see Mundy et al. 1996; Wilner et al. 2000).

The TW Hya 89 GHz continuum flux measurement agrees well with expectations 
from previously reported measurements at both higher (350~GHz) and lower 
(43~GHz) radio frequencies (Weintraub, Sandell \& Duncan 1989, 
Wilner et al. 2000, Wilner 2001).  
In addition, it is consistent with predictions 
of various disk models based on the radio observations that contain a 
population of large dust grains to account for the shallow spectral slope 
with power law index $\lesssim3$ observed in this frequency range 
(Trilling et al. 2001, Calvet et al. 2002). The good fit of the new 89 GHz 
measurement with expectations gives confidence in the reliability of the
ATCA system, and the accuracy of the absolute flux scale.

\begin{figure}
\epsscale{0.9}
\rotatebox{-90}{
\plotone{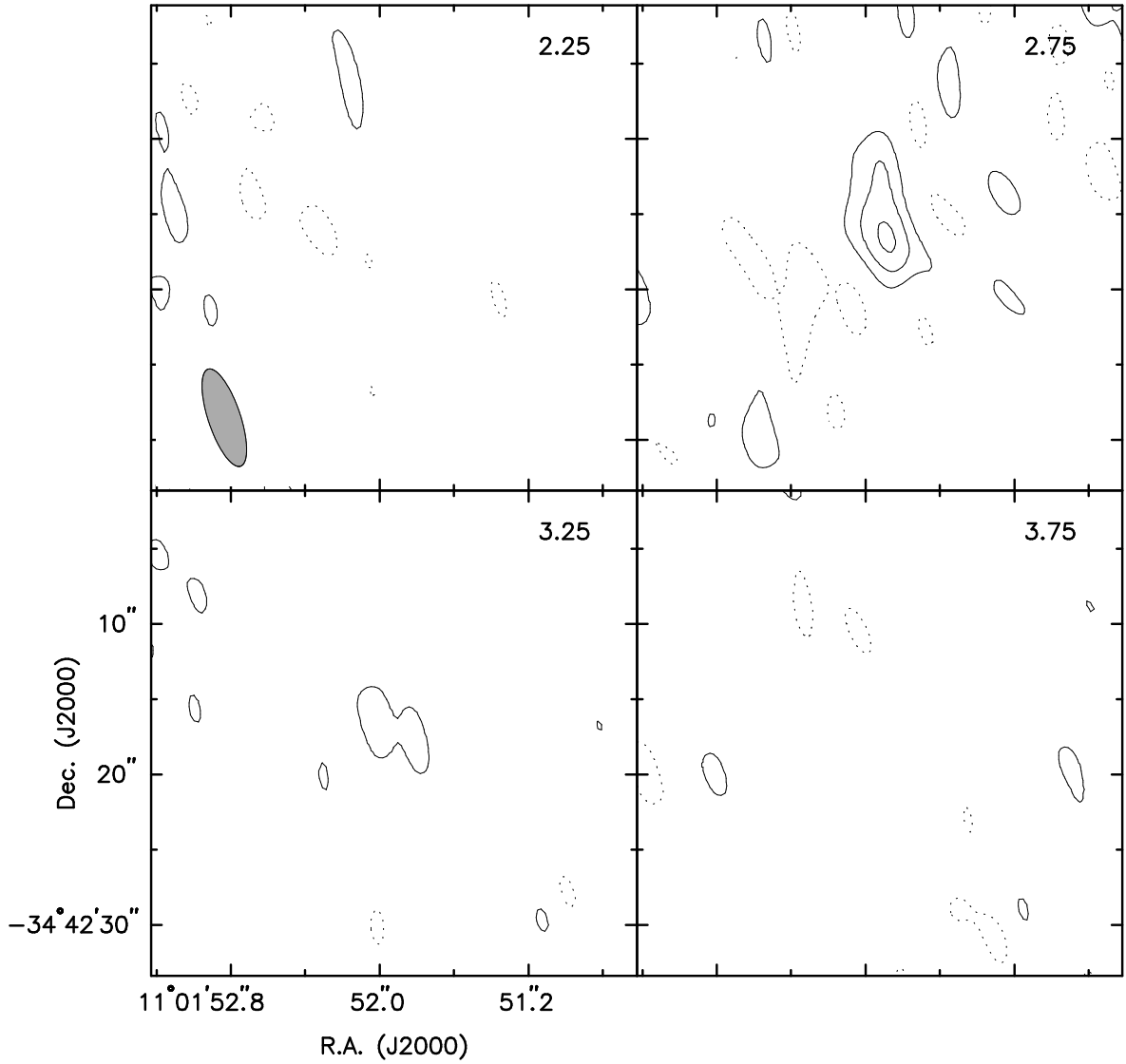}}
\caption{
Velocity channel maps (width 0.5 km~s$^{-1}$) of 
HCO$^+$ J=1-0 line emission observed from TW Hya.
The contour levels are $\pm2,4,6,...\times66$~mJy (0.7 K).
Negative contours are dotted.
The ellipse in the lower left corner of the upper left panel shows the 
$6\farcs9\times2\farcs1$ p.a. $18^{\circ}$ synthesized beam. 
}
\label{fig:tw_4chan}
\end{figure}

\begin{figure}
\epsscale{0.55}
\rotatebox{-90}{
\plotone{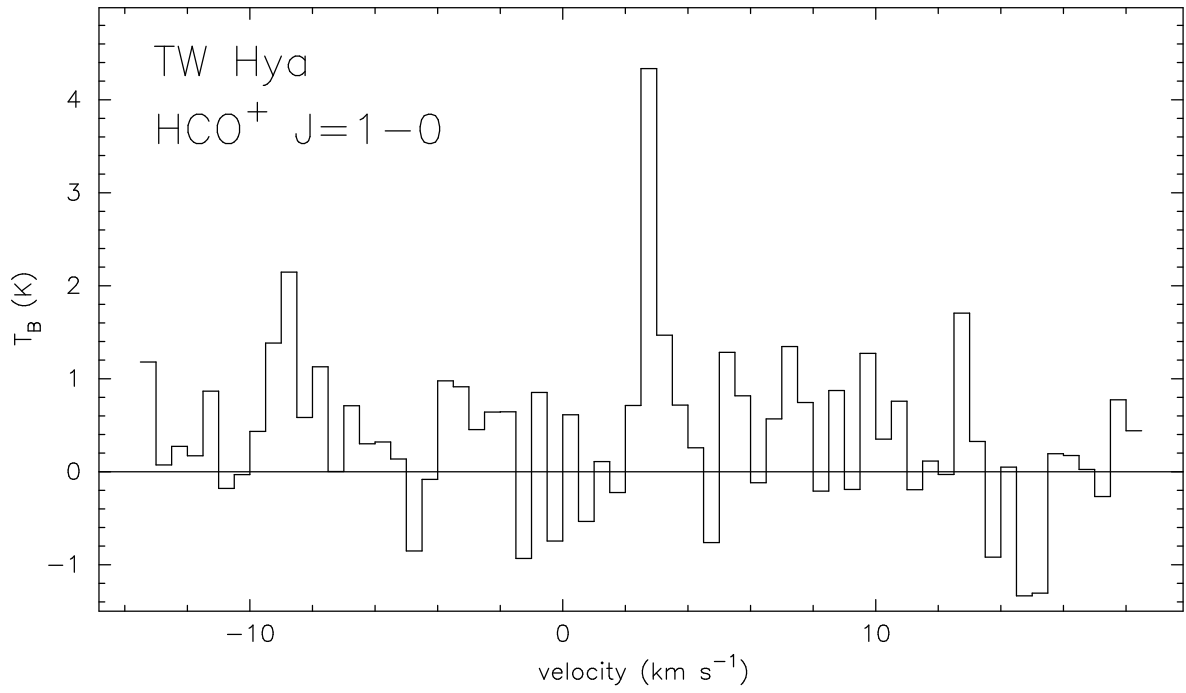}}
\caption{
Spectrum of HCO$^+$ J=1-0 line emission observed from TW Hya,
at the position of peak continuum emission, 
binned to 0.5 km~s$^{-1}$ velocity resolution, in the
$6\farcs9\times2\farcs1$ p.a. $18^{\circ}$ synthesized beam. 
}
\label{fig:tw_spec}
\end{figure}

Figure~\ref{fig:tw_4chan} shows a series of images in four 
velocity bins for the HCO$^+$ J=1--0 line emission near 
LSR velocity $\sim3$~km~s$^{-1}$.
A narrow line emission feature is clearly visible at the stellar 
position with a velocity and linewidth commensurate with previously 
reported molecular line detections from the face-on disk,
which have $V_{\rm LSR}$=2.9 km s$^{-1}$ and
$\Delta V\approx 0.6$ km s$^{-1}$. 
Figure~\ref{fig:tw_spec} shows the spectrum at the continuum position. 
The peak line flux at 0.5~km~s$^{-1}$ resolution is 0.40 Jy, 
which corresponds to a brightness temperature of 4.3~K in the 
$6\farcs9\times2\farcs1$ beam. 
The line emission from the disk is clearly spatially resolved;
a circular Gaussian fit to the visibilities in a 0.5~km~s$^{-1}$ bin
gives a full-width at half maximum size of 
$3\farcs2\pm0\farcs8$.

\subsection{HD~100546}
\begin{figure}
\epsscale{0.9}
\rotatebox{-90}{
\plotone{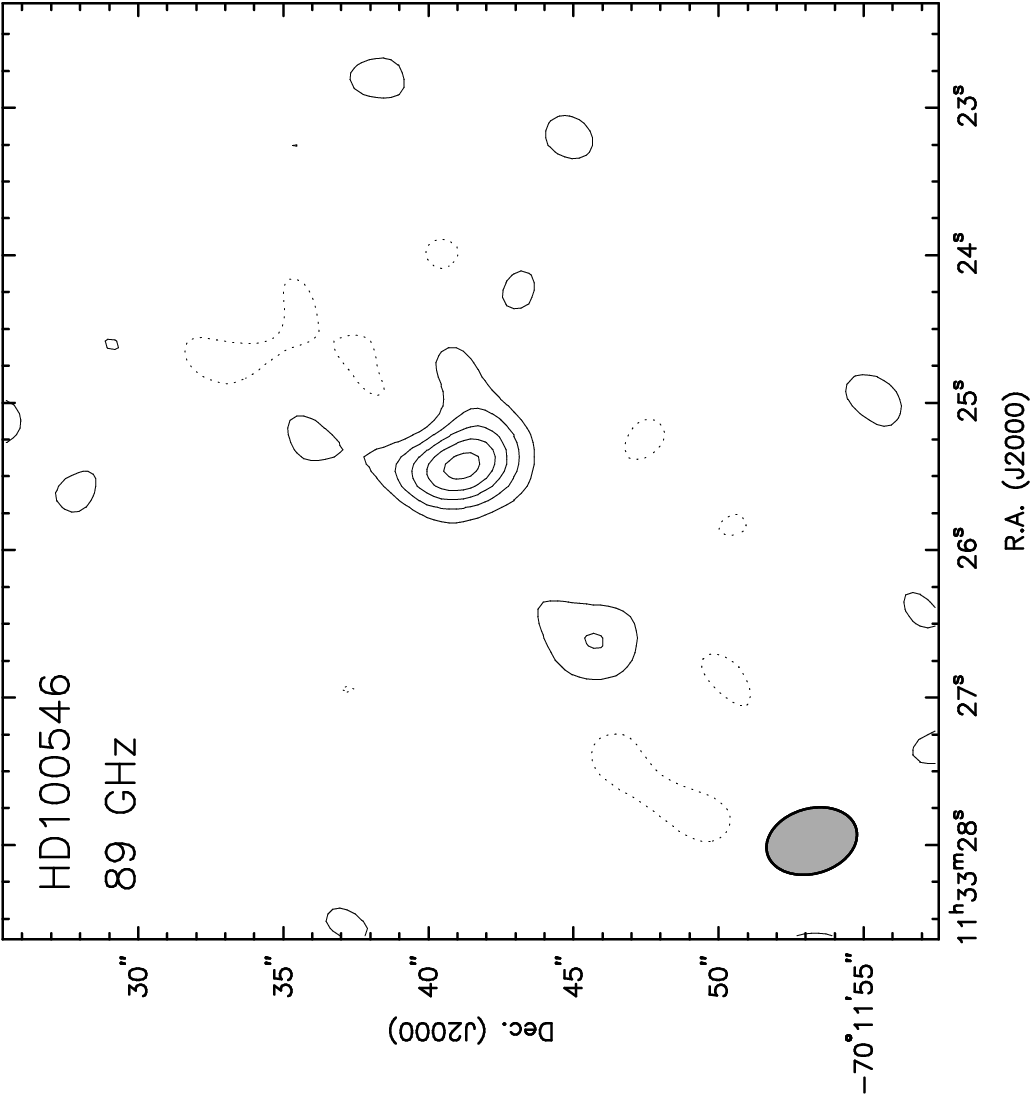}}
\caption{
HD~100546 continuum emission at 89 GHz.
The contour levels are $\pm2,4,6,...\times3.1$~mJy.
Negative contours are dotted.
The ellipse in the lower left corner shows the 
$3\farcs2\times2\farcs2$ p.a. $16^{\circ}$ synthesized beam. 
}
\label{fig:hd_cont}
\end{figure}

Figure~\ref{fig:hd_cont} shows the 89 GHz continuum emission  
detected from HD~100546. The peak position is coincident with the 
star to better than $1''$. 
A fit to the visibilities gives a point source flux of $36\pm3$ mJy 
at the star position. The residuals to this fit provide a hint 
that additional emission is present. The image shows an extra peak 
of low significance located $\sim7''$ to the southeast. This additional
peak may be related to the slight extension seen in the single dish 
1.3~mm bolometer image of Henning et al. (1998). 
Extensions along the same position angle are also visible in 
scattered light, in particular in the sensitive HST/STIS 
images of Grady et al. (2001), where nebulosity can be seen 
to extend from the northwest to the southeast.
No line emission is detected to a limit of 0.17~Jy per channel
($3\sigma$), either at the LSR velocity of the cloud DC296.2-7.9 
of 3.6~km~s$^{-1}$ (Otrupcek, Hartley \& Wang 2000),
or at the stellar LSR velocity of about -1~km~s$^{-1}$, 
which derives from the heliocentric ``astrometric'' radial velocity 
of $9\pm1$~km~s$^{-1}$ determined from Hipparcos (Madsen 2002).

\section{Discussion}
\label{sec:discussion}

\subsection{TW Hya}
The spatially resolved observations of the HCO$^+$ J=1--0 emission from 
TW Hya allow for a check of the basic paradigm that trace molecules are 
present in an extended upper layer of the disk, with overall abundances 
depleted with respect to molecular hydrogen by large factors by comparison 
with those of interstellar dark clouds 
(Aikawa \& Herbst 1999, Willacy \& Langer 2000, Aikawa et al. 2002).
In the disk midplane, which contains most of the mass, the combination 
of high densities and low temperatures results in nearly all trace
molecules sticking to grain surfaces and disappearing from the gas. 
At the disk surface, the molecules are photo-dissociated. In between, 
below the surface, is a layer that is warm and shielded from stellar 
radiation and activity where molecules survive and abundances peak.

The TW Hya disk structure and chemistry has been considered in detail 
by van Zadelhoff et al. (2001), whose model calculations suggest 
that the depletion of species like CO and HCO$^+$ results from a 
combination of photodissociation in the warm surface layers and 
freezing-out in the cold, dense parts of the disk.
In these models, the molecular emission largely originates from the 
region just below the disk surface where the HCO$^+$ abundance climbs 
to a few times $10^{-10}$. In general, the HCO$^+$ abundance follows 
the CO abundance in the disk, since its formation is primarily from 
CO reacting with H$_3^+$, with destruction by dissociative recombination 
with free electrons.  

\begin{figure}
\epsscale{0.9}
\rotatebox{-90}{
\plotone{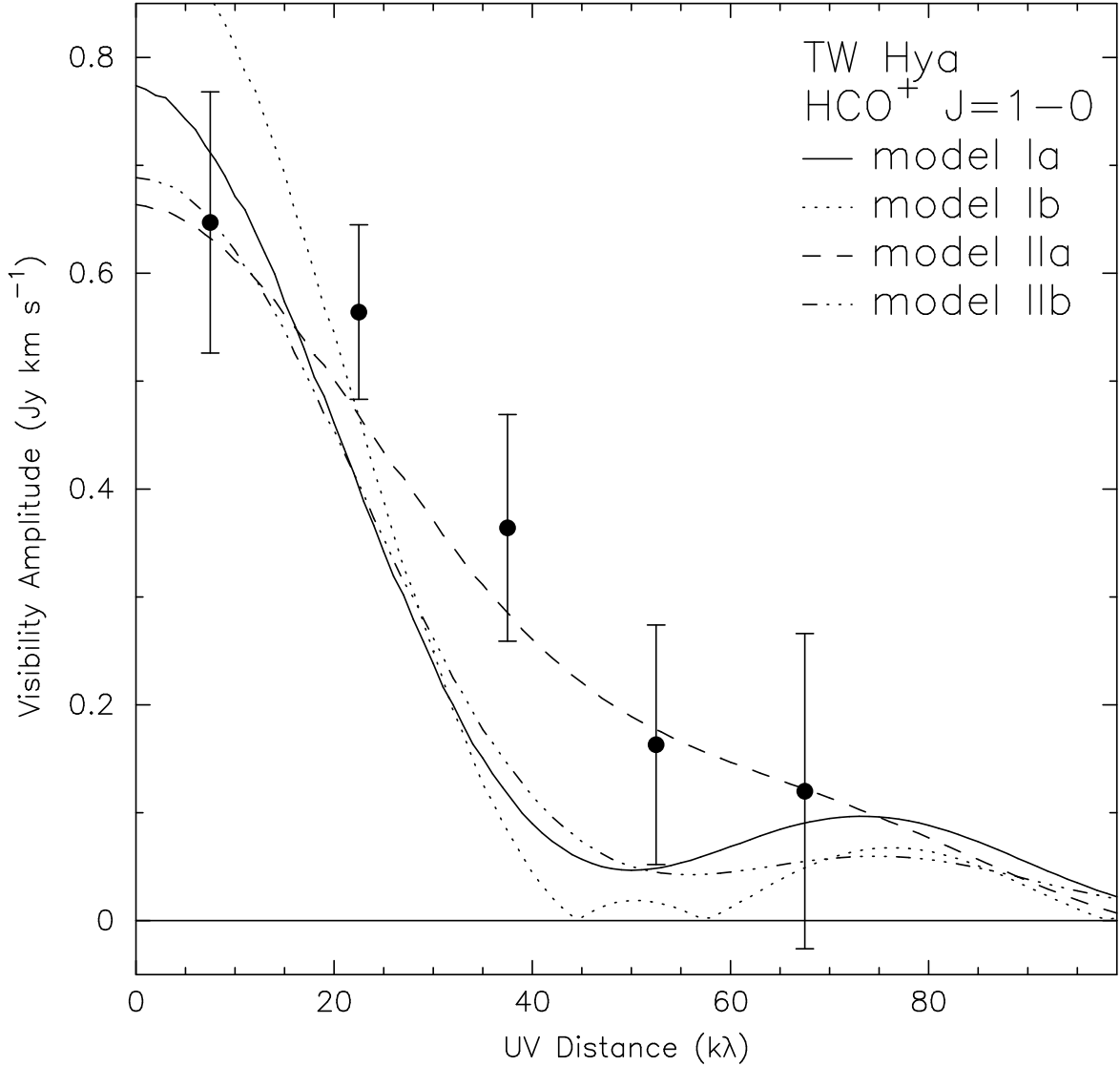}}
\caption{
Visibility amplitude of observed and modeled HCO$^+$ J=1-0 emission 
vs. baseline length, integrated over 0.5~km~s$^{-1}$ width centered at 
LSR velocity 2.75~km~s$^{-1}$, annularly averaged in 15~k$\lambda$ bins. 
The error bars represent $\pm1$ standard deviation for each bin.
The four curves show the visibility amplitudes derived from the model 
calculations of van Zadelhoff et al. (2001) that fit the single dish 
observations of higher-J HCO$^+$ transitions.
Models Ia and Ib are based on the radiatively heated disk models of
D'Alessio et al. (1999), and models IIa and IIb are based on the passive
two-layer models of Chiang \& Goldreich (1997).
For model Ia (solid line) and model IIa (dashed line), HCO$^+$ is depleted by 
an overall factor of 100 with an additional order of magnitude drop when 
the temperature falls below 20~K. For model Ib (dotted line) and 
model IIb (dash-dotted line), HCO$^+$ emission is depleted by a factor of 500. 
}
\label{fig:tw_hcop_visamp}
\end{figure}

Figure~\ref{fig:tw_hcop_visamp} shows the visibility amplitude of the 
HCO$^+$ J=1--0 emission from TW Hya as a function of baseline length; 
the falloff at longer baselines demonstrates that the emission region 
is resolved. A circular Gaussian FWHM ``size'' of $3\farcs2$ provides a 
crude description of the full line brightness distribution.  
Figure~\ref{fig:tw_hcop_visamp} includes the visibility amplitudes 
derived from four of the favored models of van Zadelhoff et al. (2001), 
which provide reasonable fits to higher-J HCO$^+$ single dish observations.
Two models are based on the radiatively heated accretion disk structure 
of D'Alessio et al. (1999), and two models are based on the passive 
two-layer description of Chiang \& Goldreich (1997).  
Two chemical scenarios are presented for each of these physical models.
The interstellar HCO$^+$ abundance is assumed to be $5\times10^{-9}$,
and in one model HCO$^+$ is depleted by a factor of 100 with an additional 
order of magnitude drop when the temperature falls below 20~K, while in
another model HCO$^+$ is depleted by a factor of 500 throughout.
Table~\ref{tab:models} summarizes the models.
Figure~\ref{fig:tw_hcop_visamp} shows that all of these models, which have 
disk masses of 0.03 $M_{\odot}$ and disk
radii of 200 AU, agree well with the ATCA observations, for 
both the size scale and the absolute intensity of the HCO$^+$ line emission.  
(Note that visibility amplitude is a positive definite quantity, 
which results in a positive bias in Figure~\ref{fig:tw_hcop_visamp}.) 

\begin{figure}
\epsscale{0.3}
\rotatebox{0}{
\plotone{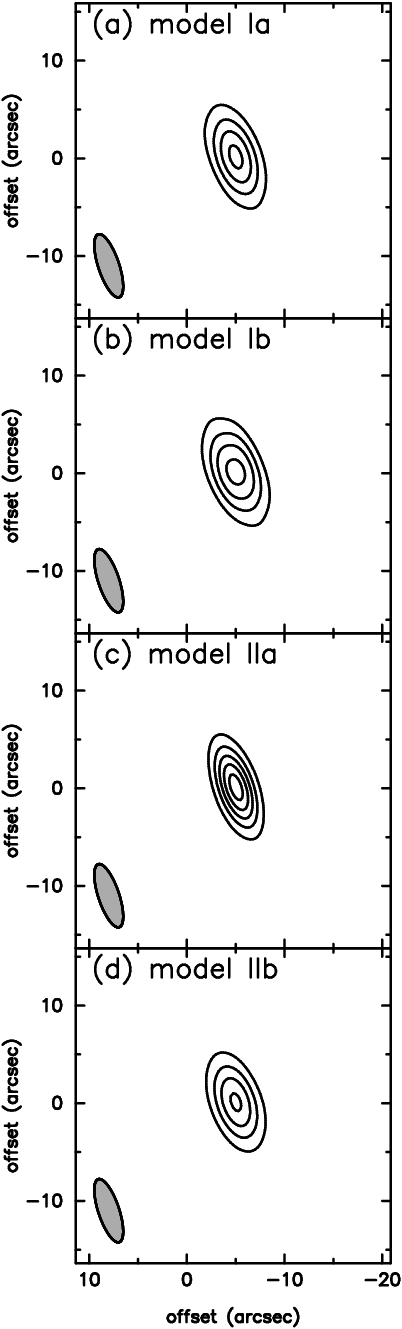}}
\caption{
Synthetic images of the HCO$^+$ J=1-0 emission for the four physical 
and chemical models whose visibility amplitude distributions 
are shown in Figure~\ref{fig:tw_hcop_visamp} 
(see~\S\ref{sec:discussion} for descriptions).
For each image, the velocity width is 0.5 km~s$^{-1}$. 
The contours levels are $\pm2,4,6,...\times66$~mJy, 
the same as in Figure~\ref{fig:tw_4chan}. 
The ellipse in the lower left corner of each panel shows the synthesized beam. 
}
\label{fig:model_images}
\end{figure}

Figure~\ref{fig:model_images} provides another view of the four models 
in the form of synthetic images. These images were made by sampling each
of the models with the same visibility distribution as obtained by the
ATCA observations, taking a 0.5~km~s$^{-1}$ bin centered on the line. 
These images may be compared with the panel in Figure~\ref{fig:tw_4chan}
for LSR velocity 2.75~km~s$^{-1}$, where the line emission peaks.
In this view, it is not easy to see that the line emission is well resolved 
spatially, but the similarity of the models in overall brightness and 
spatial extent is clearly apparent.

While none of the models of van Zadelhoff et al. (2001) were fine-tuned 
to match all of the available constraints, and none provides a unique 
best fit to the ATCA data, the high HCO$^+$ depletion factors are a robust 
feature. 
All four models are indistinguishable within the noise of the observations,
though there are differences in detail in the visibility distributions 
among the models. There is perhaps a hint that the models that produce
the more concentrated emission distributions are best.
The model curves in Figure~\ref{fig:tw_hcop_visamp} suggest that 
resolved observations with better signal-to-noise have the potential 
to discriminate between subtly different physical and chemical scenarios.

HCO$^+$ depletions of an order of magnitude or more have been inferred for 
other T-Tauri systems from millimeter imaging and analysis, in particular 
the disks around the GG Tau binary (Guilloteau, Dutrey \& Simon 1999) and 
around LkCa15 (Duvert et al. 2000, Qi 2001).
These isolated pre-main-sequence systems are located in holes of the 
Taurus dark cloud complex, which make their disks readily accessible 
without confusion. The radii of the gas disks in these systems, 
$\sim$800~AU and $\sim$600~AU, respectively, are among the largest known, 
and larger by 
factors of 3 to 4 than the $\sim200$~AU radius disk surrounding TW Hya. 
That the depletion is similarly high in all of these disks provides 
further evidence that substantial depletion must occur in the dense 
shielded midplane material, rather than simply in the colder, outer parts 
of the disks.

Since the critical density of the HCO$^+$ J=1--0 line for
collisional excitation is $\sim6\times10^4$~cm$^{-3}$, 
the detection of extended emission indicates high densities 
must be present to large radii, independent of any detailed 
physical and chemical model for the disk. 
This {\em in situ} measurement of local density confirms indications 
from scattered light models and also analysis of dust emission that 
relied on extrapolation of the inner disk density distribution.
Resolved images of multiple transitions of multiple species may be 
used to constrain the disk mass (see Dutrey, Guilloteau and Guelin 1997).

\subsection{HD~100546}
For HD~100546, the spherical radiative transfer models of 
Henning et al. (1994) that reproduce the far-infrared emission 
do not account for the observed millimeter flux, 
even with a population of ``fluffy'' dust grains, which suggests
that a circumstellar disk component must be present in the system.
The compact 89~GHz dust continuum emission detected in the 
ATCA observations provides direct evidence for this disk.
An extrapolation of the 1.3~mm flux of 0.69 Jy (Henning et al. 1998) 
to 89 GHz gives 40 mJy, for a dust opacity power law index near unity,
compared with our detected flux of $36\pm3$~mJy.
Thus, the disk component nominally accounts for a substantial fraction 
of the detected 89 GHz flux, though the uncertainties are large.
Detailed modeling of the disk properties awaits 
additional millimeter and submillimeter observations with sufficient 
angular resolution to separate the disk and envelope components.

The similar continuum fluxes of HD~100546 and TW Hya suggest that the
disk in the more distant and warmer HD~100546 system is substantially
more massive than the disk around TW Hya. Since HCO$^+$ J=1-0 emission 
is not detected from HD~100546 with comparable sensitivity to 
the TW Hya detection, it is likely that the abundance reduction is 
even more extreme. Perhaps photodissociation of CO in the disk 
atmosphere is enhanced in the environment of this more massive, hotter star.

Grady et al. (1997) suggest that much of the accreting gas seen in the 
optical along the line-of-sight to HD 100546 is associated with 
star-grazing planetesimals, such as comets or asteroids. This may explain 
both the lack of observed boundary-layer emission, which might be expected 
from accretion of mostly gaseous material, and signatures of bipolar outflow 
in forbidden line emission that is characteristic of an optically thick, 
gaseous accretion disk. This fact, together with the non-detection of 
HCO$^+$ emission and the highly processed nature of the dust around HD~100546, 
perhaps indicates that the gas is nearly fully depleted in the disk, in which 
case the planet building process may be well advanced.

\section{Conclusions}
We have made high resolution observations of the nearby young stars 
TW~Hya and HD~100546 with the ATCA using three antennas equipped at 89 GHz. 
These observations indicate the promise for future millimeter-wave 
observations of protoplanetary disks located deep in the southern sky. 
The observations of the TW Hya disk resolve emission from the 
HCO$^+$ J=1--0 line, and the measured extent and brightness support
disk models with high depletion factors inferred from previous, spatially 
unresolved, measurements of higher J transitions. The detailed modeling 
of the TW~Hya disk represents a first step towards probing the gas content 
of this important system, whose dispersal and chemistry may be similar to 
the early Solar System. 
A more complete understanding of the details of disk chemistry will
require spatially resolved images with higher sensitivity, eventually 
sampling species that probe all of the relevant chemical families.

Thermal dust continuum emission from these sources was easily detected and 
promises to be resolved with observations using longer baselines. 
The available ATCA antenna stations will offer baselines up to 1.5~km 
in length, corresponding to an angular resolution better than $0\farcs5$ 
at this frequency, which will be adequate to probe the disk surface density 
and to start to disentangle disk structure from opacity effects. 
Both TW Hya and HD~100546 will be prime protoplanetary disk targets 
for the Atacama Large Millimeter Array, the next generation millimeter 
interferometer sited in northern Chile, whose construction has 
just started and will continue for the next decade. 

\acknowledgements
We thank an anonymous referee for several suggestions that improved 
this paper.
The Australia Telescope Compact Array is part of the Australia Telescope,
which is funded by the Commonwealth of Australia for operation as a 
national facility managed by CSIRO.
Partial support for this work was provided by 
NASA Origins of Solar Systems Program Grant NAG5-11777,
by the NRAO Foreign Telescope Travel Fund Program,
by a NOVA network 2 grant and a NWO-Spinoza grant.
CMW acknowledges support of an Australian Research Council Fellowship.

\clearpage

\begin{deluxetable}{lcc}
\tablecolumns{3}
\tablewidth{0pt}
\tablecaption{Observational Parameters}
\tablehead{
\colhead{} & \colhead{TW Hya} & \colhead{HD~100546} }
\startdata
Observations: & 2002 Jun 1 & 2002 May 31  \\
~             & 2002 Aug 7 & 2002 Aug 8 \\
Min/Max baseline: & \multicolumn{2}{c}{22 to 230~meters} \\
Pointing center (J2000): & 
  $\alpha=11^{h}01^{m}52\fs31$ &  $\alpha=11^{h}33^{m}26\fs40$ \\
~ & 
  $\delta=-34^{\circ}42''17.0\farcs3$ & $\delta=-70^{\circ}11''41\farcs2$ \\
Phase calibrator: & J1147-381 & J1058-800\\
Flux calibrator: & Mars & Mars \\
Primary beam HPBW: & $37''$ & $37''$ \\
Synthesized beam HPBW: 
   & $6\farcs9\times2\farcs1$ P.A. $18^{\circ} $
   & $3\farcs2\times2\farcs2$ P.A. $16^{\circ}$ \\
K/Jy:                 & 10.6 & 21.1 \\
r.m.s. (continuum image): & 3 mJy/beam & 3 mJy/beam \\
detected flux: & $41\pm4$ mJy\tablenotemark{a} & $36\pm3$ mJy\tablenotemark{a} \\
Spectral Line Correlator: & \multicolumn{2}{c}{257 channels, 16 MHz} \\
~~~species/transition: & \multicolumn{2}{c}{HCO$^+$ J$=1-0$} \\
~~~frequency:          & \multicolumn{2}{c}{89.188518 GHz} \\
~~~center velocity: & \multicolumn{2}{c}{0 km~s$^{-1}$} \\
~~~channel spacing:  & \multicolumn{2}{c}{0.21 km~s$^{-1}$} \\
r.m.s. (line images): & 60 mJy/beam & 60 mJy/beam \\
\enddata
\tablenotetext{a}{Includes random error only; 
the systematic uncertainty in the flux scale is estimated to be $\sim20$\%.}
\label{tab:obs}
\end{deluxetable}

\clearpage

\begin{deluxetable}{llc}
\tablecolumns{3}
\tablewidth{0pt}
\tablecaption{Disk Models\tablenotemark{a}}
\tablehead{
\colhead{Model} & 
  \colhead{Physical Structure (mass 0.03 $M_{\odot}$, radius 200 AU)} & 
  \colhead{HCO$^+$ depletion} }
\startdata
Ia  & irradiated accretion disk (D'Alessio et al. 1999) &
  100$\times$ (1000$\times$ for T$<20$ K) \\
Ib  & irradiated accretion disk (D'Alessio et al. 1999) &
  500$\times$ \\
IIa & passive two-layer disk (Chiang \& Goldreich 1999) &
  100$\times$ (1000$\times$ for T$<20$ K) \\
IIb & passive two-layer disk (Chiang \& Goldreich 1999) &
  500$\times$ \\
\enddata
\tablenotetext{a}{
See the discussion by van Zadelhoff et al. (2001) 
for more details about these models and their comparable fits 
to unresolved observations of higher-J HCO$^+$ transitions.}
\label{tab:models}
\end{deluxetable}

\clearpage

\clearpage

\clearpage

\clearpage

\clearpage

\end{document}